%% file: article.tex
\documentclass[copyright,noncommercial]{eptcs}

\usepackage{amsmath,latexsym,amsfonts,amssymb,amsthm}
\usepackage{rotating,graphs,stmaryrd}

\providecommand{\event}{}

\providecommand{\firstpage}{1}

\input{macros}

\theoremstyle{plain}

\title{The Design of GP 2}

\author{Detlef Plump 
\institute{Department of Computer Science\\
           The University of York, UK}}

\def\titlerunning{The Design of GP 2}
\def\authorrunning{D. Plump}

\begin{document}

\maketitle
\thispagestyle{empty}

\begin{abstract}
This papers defines the syntax and semantics of GP 2, a revised version of the graph programming language GP. New concepts are illustrated and explained with example programs. Changes to the first version of GP include an improved type system for labels, a built-in marking mechanism for nodes and edges, a more powerful \texttt{edge} predicate for conditional rule schemata, and functions returning the indegree and outdegree of matched nodes. Moreover, the semantics of the branching and loop statement have been simplified to allow their efficient implementation.
\end{abstract}

\input{introduction}

\input{gratra}
\input{rule_schema_syntax}
\input{rule_schema_semantics}
\input{programs}

\input{sos}
\input{conclusion}

\bibliographystyle{eptcs}
\bibliography{/usr/det/Bibtex/abbr,%
              /usr/det/Bibtex/absredu.bib,%
              /usr/det/Bibtex/trs,%
              /usr/det/Bibtex/gragra,%
              /usr/det/Bibtex/graph-algorithms,%
              /usr/det/Bibtex/graphs-misc,%
              /usr/det/Bibtex/gratra-languages,%
              /usr/det/Bibtex/graph-matching,%
              /usr/det/Bibtex/proglang,%
              /usr/det/Bibtex/verification,%
              /usr/det/Bibtex/misc}

\appendix
\input{appendix}

\end{document}

%% file: macros.tex
\newtheorem*{proposition}{Proposition}

\theoremstyle{definition}
\newtheorem{definition}{Definition}
\theoremstyle{remark}
\newtheorem{example}{Example}

\newcommand{\tuple}[1]{\langle#1\rangle}

\newcommand{\ttt}{\texttt}
\newcommand{\csp}{\hspace{.2em}}

\newcommand{\mtt}{\mathtt}
\newcommand{\mrm}{\mathrm}

\renewcommand{\L}{\mathcal{L}}
\newcommand{\Z}{\mathbb{Z}}
\newcommand{\B}{\mathbb{B}}
\newcommand{\C}{\mathrm{Char}}

\newcommand{\Sem}[1]{\llbracket{#1}\rrbracket}

\newcommand{\G}{\mathcal{G}}
\newcommand{\R}{\mathcal{R}}

\newcommand{\x}{\mathtt{x}}

\newcommand{\failrm}{\mathrm{fail}}
\newcommand{\failtt}{\mathtt{fail}}
\newcommand{\skiptt}{\mathtt{skip}}

\newcommand{\grey}{\graphnodecolour{.75}}
\newcommand{\dashed}{\graphlinedash{4 2}}

\newcommand{\nodegraph}[1]{%
 \graphlinewidth{.01}
 \graphlinecolour{0}
 \graphnodecolour{1}
 \graphnodesize{.4}
 \begin{graph}(.4,.2)
  \roundnode{n}(.2,.1)\autonodetext{n}{#1}
 \end{graph}
}

\newcommand{\dder}{\Rightarrow}

\newcommand{\DSto}{\mathop{\to}\limits}
\newcommand{\DSdder}{\mathop{\Rightarrow}\limits}

\newcommand{\ifte}[3]{\mathtt{if}\ #1\ \mathtt{then}\ #2\ \mathtt{else}\ #3}
\newcommand{\ift}[2]{\mathtt{if}\ #1\ \mathtt{then}\ #2}
\newcommand{\tryt}[2]{\mathtt{try}\ #1\ \mathtt{then}\ #2}
\newcommand{\tryte}[3]{\mathtt{try}\ #1\ \mathtt{then}\ #2\ \mathtt{else}\ #3}

\newcommand{\fail}{\mathrm{fail}}

%% file: introduction.tex
\section{Introduction}
\label{sec:introduction}

GP is an experimental nondeterministic programming language for high-level problem solving in the domain of graphs. The language is based on conditional rule schemata for graph transformation and has a simple syntax and semantics, to facilitate both understanding by programmers and formal reasoning on programs. The original version of GP (also referred to as GP 1 from now on) is defined in \cite{Plump09a,Plump-Steinert10a} and its protoype implementation is described in \cite{Manning-Plump08b}.

Motivated by case studies in GP programming, the following changes and extensions feature in GP~2:

\begin{itemize}
\item There are new types \texttt{atom} and \texttt{list}, the former representing the union of integers and character strings, the latter lists of atoms. Variables of these types can be declared in rule schemata.
\item Rule schemata can \emph{mark}\/ nodes and edges graphically.
\item Conditional rule schemata can check, by means of the \texttt{edge} predicate, whether there exists an edge with a particular label between two matched nodes.
\item The indegree or outdegree of a matched node can be accessed and used in the labels or in the condition of a rule schema.
\item The if-then-else statement of GP 1 is complemented by a try-then-else command whose then-part is executed on the graph resulting from the try-part. Also, a new \texttt{or} command provides explicit nondeterministic choice between subprograms.
\item Failure in evaluating the condition of a branching statement or the body of a loop does no longer enforce backtracking, in order to allow an efficient implementation of branching and looping.
\end{itemize}

The rest of this paper is organised as follows. In Section \ref{sec:newgratra}, the graph transformation approach underlying GP is briefly reviewed, viz.\ the double-pushout approach with relabelling. Section \ref{sec:rule_schema_syntax} introduces conditional rule schemata, the building blocks of GP programs. The semantics of conditional rule schemata is defined in Section \ref{sec:rule_schema_semantics}. In Section \ref{sec:programs}, new features of GP 2 are demonstrated and explained by example programs. A formal operational semantics for GP 2 is presented and discussed in Section \ref{sec:sos}. Section \ref{sec:conclusion} concludes by summarising GP's revision and addressing topics for future work. The Appendix lists the inference rules of the semantics of Section \ref{sec:sos}.

%% file: gratra.tex
\section{Graphs and Graph Transformation}
\label{sec:newgratra}

Graph transformation in GP is based on the double-pushout approach with relabelling \cite{Habel-Plump02c}. This framework deals with partially labelled graphs whose definition is recalled below. In this section, we treat the label alphabet as a parameter because in subsequent sections we need different alphabets: graphs in rule schemata are labelled with expressions while graphs on which GP programs operate (also referred to as host graphs) are labelled with lists composed of integers and strings. 

A \emph{graph}\/ over a label alphabet $\mathcal{C}$ is a system $G=(V_G,E_G,s_G,t_G,l_G,m_G)$, where $V_G$ and $E_G$ are finite sets of \emph{nodes} (or \emph{vertices}) and \emph{edges}, $s_G,t_G\colon E_G\rightarrow V_G$ are the \emph{source} and \emph{target} functions for edges, $l_G\colon V_G\to \mathcal{C}$ is the partial node labelling function and $m_G\colon E_G\to \mathcal{C}$ is the (total) edge labelling function. Given a node $v$, we write $l_G(v) = \mathop{\perp}$ to express that $l_G(v)$ is undefined. Graph $G$ is \emph{totally labelled}\/ if $l_G$ is a total function. We write $\G(\mathcal{C}_{\bot})$ and $\G(\mathcal{C})$ for the class of graphs resp.\ totally labelled graphs over $\mathcal{C}$.

Unlabelled nodes will occur only in the interfaces of rules and are necessary in the double-pushout approach to relabel nodes. There is no need to relabel edges as they can always be deleted and reinserted with different labels.

A \emph{graph morphism} $g\colon G\rightarrow H$ between graphs $G,H$\/ in $\G(\mathcal{C}_{\bot})$ consists of two functions $g_V\colon V_G\rightarrow V_H$ and $g_E\colon E_G\rightarrow E_H$\/ that preserve sources, targets and labels; that is, $s_H\circ g_E=g_V\circ s_G$, $t_H\circ g_E=g_V\circ t_G$, $m_H \circ g_E = m_G$, and $l_H(g(v))=l_G(v)$ for all $v$ such that $l_G(v) \neq \mathop{\perp}$. Morphism $g$ is an \emph{inclusion} if $g(x)=x$ for all nodes and edges $x$. It is \emph{injective}\/ (\emph{surjective}) if $g_V$ and $g_E$ are injective (surjective). It is an \emph{isomorphism} if it is injective, surjective and satisfies $l_H(g_V(v)) = \bot$ for all nodes $v$ with $l_G(v) = \bot$. In this case $G$ and $H$\/ are \emph{isomorphic}, which is denoted by $G\cong H$.

A \emph{rule} $r = \tuple{L \gets K \to R}$ consists of two inclusions $K \to L$\/ and $K \to R$\/ such that $L,R$\/ are graphs in $\G(\mathcal{C})$ and $K$, the \emph{interface} of $r$, is a graph in $\G(\mathcal{C}_{\bot})$. Intuitively, an application of $r$ to a graph will remove the items in $L - K$, preserve $K$, add the items in $R - K$, and relabel the unlabelled nodes in $K$. 

\begin{definition}[Rule application]
\label{def:rule_application}
Let $r = \tuple{L \gets K \to R}$ be a rule, $G$\/ a graph in $\G(\mathcal{C})$, and $g\colon L \to G$\/ an injective graph morphism satisfying the \emph{dangling condition}: no node in $g(L)-g(K)$ is incident to an edge in $G-g(L)$. We write $G \dder_{r,g} H$\/ if $H$\/ is isomorphic to the graph that is constructed from $G$\/ as follows: 
\begin{enumerate}
\item Remove all nodes and edges in $g(L)-g(K)$, obtaining a graph $D$.
\item Add disjointly to $D$ all nodes and edges from $R-K$, keeping their labels. For $e \in E_R-E_K$, $s_H(e)$ is $s_R(e)$ if $s_R(e) \in V_R-V_K$, otherwise $g_V(s_R(e))$. Targets are defined analogously.  
\item For each unlabelled node $v$ in $K$, $l_H(g_V(v))$ becomes $l_R(v)$. 
\end{enumerate}
\end{definition}

\begin{figure}[htb]
\begin{center}
\input{Graphs/direct_derivation.graph}
\end{center}
 
\caption{A rule application \label{fig:direct_derivation}}
\end{figure}
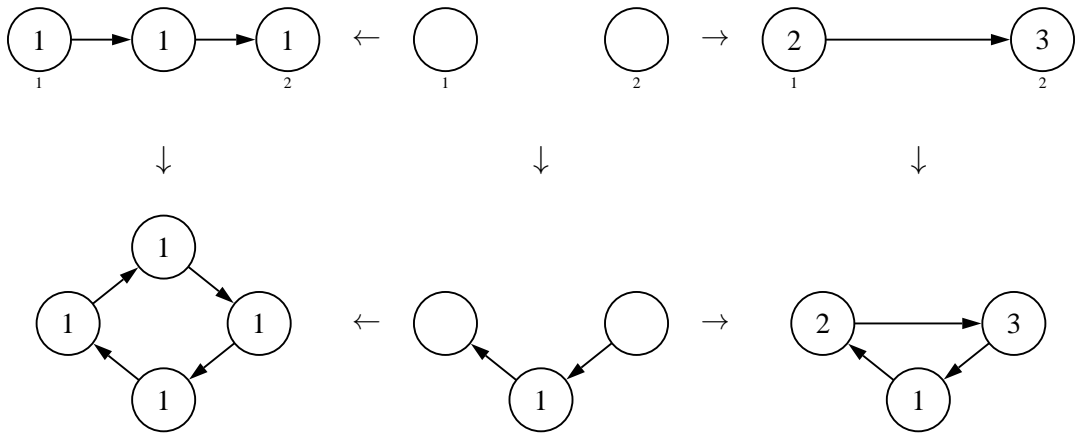

Figure \ref{fig:direct_derivation} shows an example of a rule application. The rule in the upper row is applied to the left graph of the lower row, resulting in the right graph of the lower row. (For simplicity, we assume that all edge labels are the same and hence omit them.) The node identifiers 1 and 2 in the rule specify the inclusions of the interface. The middle graph of the lower row is obtained from graph $D$\/ of Definition \ref{def:rule_application} by making all nodes unlabelled that are images of unlabelled nodes in $K$. Then the diagram represents a double-pushout in the category of graphs over $\mathcal{C}_{\bot}$ (see \cite{Habel-Plump02c}).

%% file: Graphs/direct_derivation.graph
   \setlength{\unitlength}{0.5in}
   \graphnodesize{0.68}
   \grapharrowlength{0.2}
   \grapharrowwidth{0.6}
   \autodistance{1.0}
   \graphnodecolour{1}

\begin{tabular}{c}
\begin{graph}(3.4,1.5)(0,.95)
 \roundnode{n1}(.4,1.7)\autonodetext{n1}{1}\autonodetext{n1}[s]{\tiny 1}
 \roundnode{n2}(1.7,1.7)\autonodetext{n2}{1}
 \roundnode{n3}(3,1.7)\autonodetext{n3}{1}\autonodetext{n3}[s]{\tiny 2}
 \diredge{n1}{n2}
 \diredge{n2}{n3}
\end{graph}
\begin{graph}(.7,1.5)
 \freetext(.35,.75){$\gets$}
\end{graph}
\begin{graph}(2.8,1.5)(0,.95)
 \roundnode{n1}(.4,1.7)\autonodetext{n1}[s]{\tiny 1}
 \roundnode{n3}(2.4,1.7)\autonodetext{n3}[s]{\tiny 2}
\end{graph}
\begin{graph}(.7,1.5)
 \freetext(.35,.75){$\to$}
\end{graph}
\begin{graph}(3.4,1.5)(0,.95)
 \roundnode{n1}(.4,1.7)\autonodetext{n1}{2}\autonodetext{n1}[s]{\tiny 1}
 \roundnode{n3}(3,1.7)\autonodetext{n3}{3}\autonodetext{n3}[s]{\tiny 2}
 \diredge{n1}{n3}
\end{graph}
\\
\begin{graph}(3.4,.8)
 \freetext(1.7,.4){\large $\downarrow$}
\end{graph}
\begin{graph}(.7,.8)
\end{graph}
\begin{graph}(2.8,.8)
 \freetext(1.4,.4){\large $\downarrow$}
\end{graph}
\begin{graph}(.7,.8)
\end{graph}
\begin{graph}(3.4,.8)
 \freetext(1.7,.4){\large $\downarrow$}
\end{graph}
\\
\begin{graph}(3.4,2.4)(.4,0)
 \roundnode{n1}(1.1,1.2)\autonodetext{n1}{1}
 \roundnode{n2}(2.1,2)\autonodetext{n2}{1}
 \roundnode{n3}(3.1,1.2)\autonodetext{n3}{1}
 \roundnode{n4}(2.1,.4)\autonodetext{n4}{1}
 \diredge{n1}{n2}
 \diredge{n2}{n3}
 \diredge{n3}{n4}
 \diredge{n4}{n1}
\end{graph}
\begin{graph}(.7,2.4)
 \freetext(.35,1.2){$\gets$}
\end{graph}
\begin{graph}(2.8,2.4)(.7,0)
 \roundnode{n1}(1.1,1.2)
 \roundnode{n3}(3.1,1.2)
 \roundnode{n4}(2.1,.4)\autonodetext{n4}{1}
 \diredge{n3}{n4}
 \diredge{n4}{n1}
\end{graph}
\begin{graph}(.7,2.4)
 \freetext(.35,1.2){$\to$}
\end{graph}
\begin{graph}(3.4,2.4)(.4,0)
 \roundnode{n1}(1.1,1.2)\autonodetext{n1}{2}
 \roundnode{n3}(3.1,1.2)\autonodetext{n3}{3}
 \roundnode{n4}(2.1,.4)\autonodetext{n4}{1}
 \diredge{n1}{n3}
 \diredge{n3}{n4}
 \diredge{n4}{n1}
\end{graph}
\end{tabular}

%% file: rule_schema_syntax.tex
\section{Syntax of Rule Schemata}
\label{sec:rule_schema_syntax}

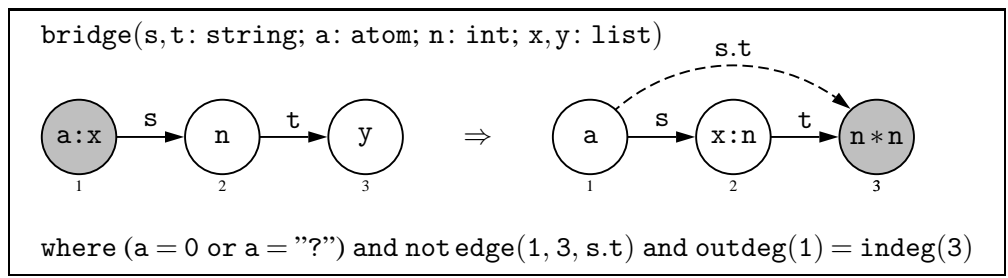
\begin{figure}[htb]
 \begin{center}
  \input{Graphs/condruleschema.graph}
 \end{center}
\caption{Declaration of a conditional rule schema}\label{fig:condruleschema}
\end{figure}

Conditional rule schemata are the principal programming construct in GP. Figure \ref{fig:condruleschema} shows an (artificial) example for the declaration of a rule schema containing some of the new features of GP 2. 

Besides the types \texttt{int} and \texttt{string} of GP 1, there are the new types \texttt{atom} and \texttt{list}. Type \texttt{atom} is the union of \texttt{int} and \texttt{string}, and \texttt{list} is the type of a (possibly empty) list of atoms. Given lists $\mtt{x}$ and $\mtt{y}$, we write \texttt{x:y} for the concatenation of $\mtt{x}$ and $\mtt{y}$. The colon replaces the underscore '\texttt{\_}' of GP 1 for better readability. Also, the empty list \texttt{empty} is now allowed (not to be confused with the empty character string ''''). When drawing graphs, we represent the empty list by omitting the word \texttt{empty}. (Confusion with unlabelled nodes is not possible as long as we consider graphs on the left or right of a rule schema, or host graphs. This is because these graphs are totally labelled.) 

\begin{figure}[htb]
\begin{center}
 \input{Graphs/subtypes.graph}
\end{center}
\caption{Subtype hierarchy for lists}\label{fig:subtypes}
\end{figure}
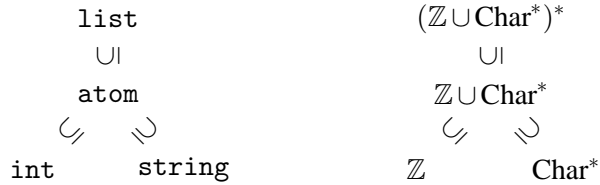

We identify lists of length one with their contents and hence get the syntactic and semantic \emph{subtype}\/ relationships shown in Figure \ref{fig:subtypes}. This is why we can form list expressions such as \texttt{a:x} and \texttt{x:n} in Figure \ref{fig:condruleschema}, where $\mtt{x}$ is a list, $\mtt{a}$ an atom and $\mtt{n}$ an integer. For the same reason, equations in the condition such as $\mtt{a=0}$ or $\mtt{a=\text{''\texttt{?}''}}$ can compare expressions of arbitrary list subtypes.

Expressions in the left-hand side of a rule schema need no longer be constants or variables. Composite expressions such as \texttt{a:x} in Figure \ref{fig:condruleschema} are allowed if there is no ambiguity in matching individual variables with values in host graph labels. Similarly, the new dot operator '\texttt{.}' for string concatenation can be used in left-hand labels. (The exact condition for left-hand expressions is given in Definition \ref{def:simple_label}.)

The new functions \texttt{indeg} and \texttt{outdeg} access the indegree resp.\ outdegree of a left-hand node in the host graph. These operators may occur in the labels and the condition of a rule schema. 
Moreover, the binary \texttt{edge} predicate of GP 1 has now an optional third argument specifying the label of a possible edge between the given nodes. For example, the subcondition $\mtt{not\, edge(1,\,3,\, s.t)}$ in  Figure \ref{fig:condruleschema} demands that there must not be an edge from node \texttt{1} to node \texttt{3} with label \texttt{s.t} (where the strings denoted by \texttt{s} and \texttt{t} are determined by matching the left-hand graph).

Finally, GP 2 allows to mark nodes and edges graphically. For example, the outermost nodes in Figure \ref{fig:condruleschema} are marked by a grey shading, and the dashed arrow between nodes 1 and 3 in the right graph represents a marked edge. Marking is formalised below by defining labels as pairs of lists and boolean values, where a boolean value indicates whether a node or edge is marked or not.

Figure \ref{fig:abstract_label_syntax} and Figure \ref{fig:abstract_condition_syntax} give grammars in Extended Backus-Naur Form defining the abstract syntax of the labels and the condition of a rule schema. These grammars are ambiguous; in examples we use parentheses to disambiguate expressions if necessary. In the next section, the abstract syntax is used in defining the semantics of rule schemata.

\begin{figure}[htb] 
\begin{center}
\renewcommand{\arraystretch}{1.2}
\begin{tabular}{lcl}
Integer & ::= & Digit \{Digit\} $\mid$ IVariable $\mid$ '$-$' Integer $\mid$ Integer ArithOp Integer $\mid$ \\
&& (\texttt{indeg} $\mid$ \texttt{outdeg}) '(' Node ')' \\
ArithOp & ::= & '\texttt{+}' $\mid$ '\texttt{-}' $\mid$ '$\ast$' $\mid$ '\texttt{/}' \\
String & ::= & '\,''\,' \{Char\} '\,''\,' $\mid$ SVariable $\mid$ String '.' String \\
Atom & ::= & Integer $\mid$ String $\mid$ AVariable \\
List & ::= & \texttt{empty} $\mid$ Atom $\mid$ LVariable $\mid$ List ':' List \\
Label & ::= & List Mark \\
Mark & ::= & \texttt{true} $\mid$  \texttt{false} 
\end{tabular}
\end{center}
\caption{Abstract syntax of rule schema labels\label{fig:abstract_label_syntax}
}
\end{figure}

The grammar in Figure \ref{fig:abstract_label_syntax} defines four syntactic categories of expressions which can occur in a rule schema: Integer, String, Atom and List, where Integer and String are subsets of Atom which in turn is a subset of List. We assume that Node is the set of node identifiers occurring in the rule schema, which must be the same for the left and the right graph ($\{1,2,3\}$ in  Figure \ref{fig:condruleschema}). Moreover, IVariable, SVariable, AVariable and LVariable are the sets of variables of type \texttt{int}, \texttt{string}, \texttt{atom} and \texttt{list} that occur in the rule schema. These categories are disjoint since each variable must be declared with a unique type (see Figure \ref{fig:condruleschema}). The mark components of labels are represented graphically rather than textually. 

The values of variables at execution time are determined by graph matching, hence we require that expressions in the left graph of a rule schema must have a simple shape.

\begin{samepage}
\begin{definition}[Simple list]
\label{def:simple_label}
An expression $e \in \mrm{List}$ is \emph{simple}\/ if
\begin{enumerate}
\item[(1)]  $e$\/ contains no arithmetic operators,
\item[(2)] $e$\/ contains at most one occurrence of a list variable, and
\item[(3)] each occurrence of a string expression in $e$\/ contains at  most one occurrence of a string variable.
\end {enumerate}
\end{definition}
\end{samepage}
For example, given the variable declarations of Figure \ref{fig:condruleschema}, \texttt{a:x} and ''\texttt{no}''\texttt{.s:y:t} are simple expressions whereas \texttt{x:y} and \texttt{s.t} are not simple.

The syntax of a rule schema condition is defined by the grammar in Figure \ref{fig:abstract_condition_syntax}. New features are the predicates \texttt{int}, \texttt{string} and \texttt{atom}, which allow to check whether an expression belongs to a subtype of \texttt{list}, and equations between arbitrary list expressions. Also, the \texttt{edge} predicate can have a third parameter specifying the list component of an edge label. 

\begin{figure}[htb]
\renewcommand{\arraystretch}{1.2}
\begin{center}
\begin{tabular}{lcl}
Condition & ::= & Type '(' List ')' $\mid$ List ('\texttt{=}' $\mid$ '\texttt{!=}') List $\mid$ \\
&& Integer RelOp Integer $\mid$ \\
&& \texttt{edge} '(' Node ',' Node [',' List] ')' $\mid$ \\
&& \texttt{not} Condition $\mid$ Condition (\texttt{and} $\mid$ \texttt{or}) Condition \\
Type & ::= & \texttt{int} $\mid$ \texttt{string} $\mid$ \texttt{atom} \\
RelOp & ::= & '\texttt{>}' $\mid$ '\texttt{>=}' $\mid$ '\texttt{<}' $\mid$ '\texttt{<=}' 
\end{tabular}
\end{center}
\caption{Abstract syntax of rule-schema condition \label{fig:abstract_condition_syntax}}
\end{figure}

\begin{definition}[Conditional rule schema]
\label{def:condruleschema}
A \emph{rule schema} $\tuple{L \gets K \to R}$ consists of two inclusions $K \to L$\/ and $K \to R$\/ such that $L,R$\/ are graphs in $\G(\mrm{Label})$ and $K$\/ consists of unlabelled nodes only. We require that all list expressions in $L$\/ are simple and that all variables occurring in $R$\/ also occur in $L$. A \emph{conditional}\/ rule schema $\tuple{L \gets K \to R,\, c}$ consists of a rule schema $\tuple{L \gets K \to R}$ and a condition $c \in \mrm{Condition}$ such that all variables occurring in $c$\/ also occur in $L$. 
\end{definition}

When a conditional rule schema is declared, as in Figure \ref{fig:condruleschema}, graph $K$\/ is implicitly represented by the node identifiers in $L$\/ and $R$\/ (which much coincide). Hence nodes without identifiers in $L$\/ are to be deleted and nodes without identifiers in $R$\/ are to be created.

The requirement that all variables in $R$\/ must also occur in $L$\/ ensures that for a given match of $L$\/ in a host graph, applying $r$ produces a unique graph (up to isomorphism). Similarly, the evaluation of $c$ has a unique result if all its variables occur in $L$.

%% file: Graphs/condruleschema.graph
   \setlength{\unitlength}{0.5in}
   \graphnodesize{0.8}
   \grapharrowlength{0.2}
   \grapharrowwidth{0.6}
   \autodistance{1.0}
   \graphnodecolour{1}
\fbox{
\begin{tabular}{l}
$\mtt{bridge (s,t\colon string;\; a\colon atom;\; n\colon int;\; x,y\colon list)}$
\\
\begin{graph}(3.8,1.74)(.2,.83)
 \roundnode{n1}(.6,1.7)[\grey]
  \autonodetext{n1}{\texttt{a:x}}\autonodetext{n1}[s]{\tiny 1}
 \roundnode{n2}(2.1,1.7)
  \autonodetext{n2}{\texttt{n}}\autonodetext{n2}[s]{\tiny 2}
 \roundnode{n3}(3.6,1.7)
  \autonodetext{n3}{\texttt{y}}\autonodetext{n3}[s]{\tiny 3}
 \diredge{n1}{n2}\bowtext{n1}{n2}{.12}{\texttt{s}}
 \diredge{n2}{n3}\bowtext{n2}{n3}{.12}{\texttt{t}}
\end{graph}
\begin{graph}(1.4,1.74)(0,-.1)
 \freetext(.7,.75){$\dder$}
\end{graph}
\begin{graph}(3.8,1.74)(.2,.83)
 \roundnode{n1}(.6,1.7)
  \autonodetext{n1}{\texttt{a}}\autonodetext{n1}[s]{\tiny 1}
  \roundnode{n2}(2.1,1.7)
  \autonodetext{n2}{\texttt{x:n}}\autonodetext{n2}[s]{\tiny 2}
 \roundnode{n3}(3.6,1.7)[\grey]\autonodetext{n3}[s]{\tiny 3}
  \autonodetext{n3}{$\mtt{n \ast n}$}\autonodetext{n3}[s]{\tiny 3}
 \dirbow{n1}{n3}{.25}[\dashed]\bowtext{n1}{n3}{.31}{$\mathtt{s.t}$}
 \diredge{n1}{n2}\bowtext{n1}{n2}{.12}{\texttt{s}}
 \diredge{n2}{n3}\bowtext{n2}{n3}{.12}{\texttt{t}}
\end{graph}
\\
\texttt{where} ($\mtt{a=0}$ \texttt{or} $\mtt{a=\text{''\texttt{?}''}}$) \texttt{and} $\mtt{not\, edge(1,\,3,\, s.t)}$  \texttt{and} $\mtt{outdeg(1) = indeg(3)}$
\end{tabular}
}

%% file: Graphs/subtypes.graph
   \graphnodesize{0.3}
   \graphlinewidth{0.01}
   \grapharrowlength{0.1}
   \grapharrowwidth{0.5}

\begin{center}
\begin{tabular}{ccc}
\begin{graph}(3.25,3)
\opaquetextfalse
\freetext(1.5,2.5){\texttt{list}}
\freetext(1.5,1.5){\texttt{atom}}
\freetext(.5,.5){\texttt{int}}
\freetext(2.5,.5){\texttt{string}}
\freetext(1.5,2){\begin{turn}{90} \mbox{$\subseteq$} \end{turn}}
\freetext(1,1){\begin{turn}{45} \mbox{$\subseteq$} \end{turn}}
\freetext(2,1){\begin{turn}{-45} \mbox{$\supseteq$} \end{turn}}
\end{graph}
& \hspace{1cm} &
\begin{graph}(3.25,3)
\opaquetextfalse
\freetext(1.5,2.5){$(\Z \cup \C^*)^*$}
\freetext(1.5,1.5){$\Z \cup \C^*$}
\freetext(.5,.5){$\Z$}
\freetext(2.5,.5){$\C^*$}
\freetext(1.5,2){\begin{turn}{90} \mbox{$\subseteq$} \end{turn}}
\freetext(1,1){\begin{turn}{45} \mbox{$\subseteq$} \end{turn}}
\freetext(2,1){\begin{turn}{-45} \mbox{$\supseteq$} \end{turn}}
\end{graph}
\end{tabular}
\end{center}

%% file: rule_schema_semantics.tex
\section{Semantics of Rule Schemata}
\label{sec:rule_schema_semantics}

While the left and right graph of a rule schema are labelled with elements from the syntactic category Label, host graphs are labelled with values from the following semantic domain $ \L$:
\[ \L = (\Z \cup \C^*)^* \times \B \]
where $ \B = \{\mrm{true},\mrm{false}\}$. Hence semantic labels are sequences consisting of integers and character strings\footnote{We assume that Char is a fixed set of characters.}, paired with boolean values. As in the case of syntactic labels, the individual elements of a sequence are separated by colons, the empty sequence is represented by ``white space'', and the boolean value true is represented graphically by shading resp.\ dashed lines.

The application of a rule schema $r$ with condition $c$ to a graph $G$ in $\G(\L)$ proceeds roughly as follows:
\begin{enumerate}
\item Match the left graph $L$ of $r$ with a subgraph of $G$, ignoring labels, by means of a premorphism $g\colon L \to G$.
\item Check whether there is an assignment $\alpha$ of values to variables such that after evaluating the expressions in $L$, $g$\/ is label-preserving.
\item Check whether the condition $c$ evaluates to true.
\item Apply the rule $r^{g,\alpha}$, obtained from $r$\/ by evaluating all expressions in the left and right graph, to $G$.
\end{enumerate}

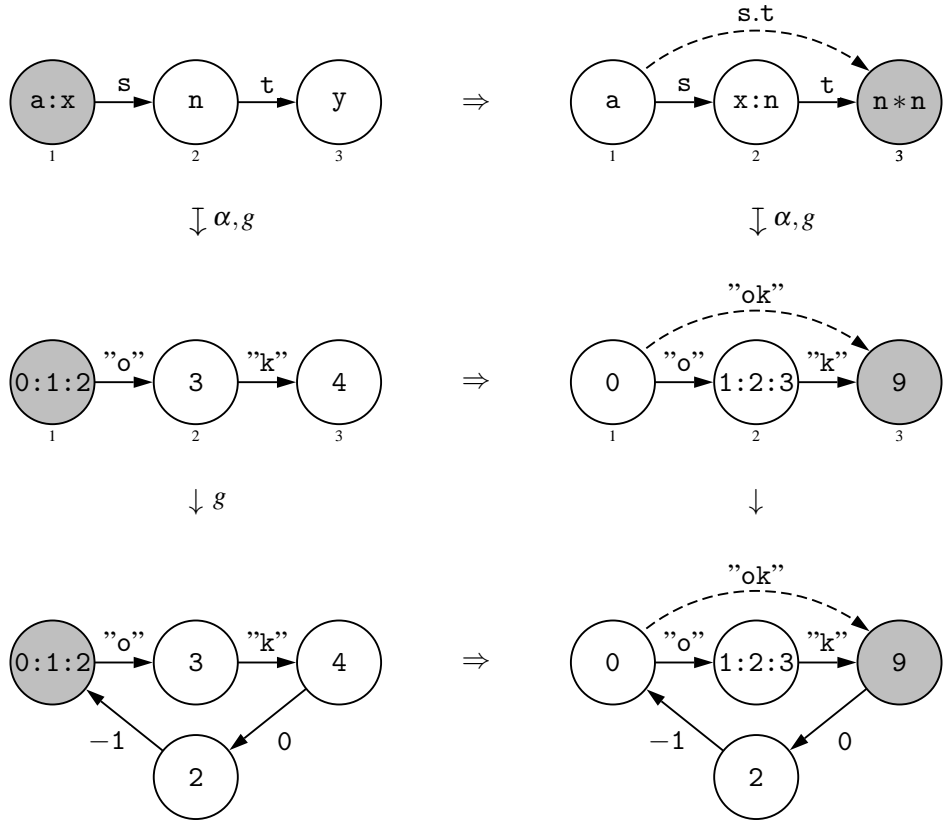
\begin{figure}[htb]
 \begin{center}
  \input{Graphs/condruleschema_appl.graph}
 \end{center}
\caption{An application of the rule schema \texttt{bridge} of Figure \ref{fig:condruleschema}}\label{fig:condruleschema_appl}
\end{figure}

For example, Figure \ref{fig:condruleschema_appl} shows an application of the rule schema \texttt{bridge} of Figure \ref{fig:condruleschema}. The upper half of the diagram represents the instantiation of \texttt{bridge} according to premorphism $g$ and the following assignment $\alpha$: $\mtt{a} \mapsto \mtt{0}$, $\mtt{x} \mapsto \mtt{1:2}$, $\mtt{n} \mapsto 3$, $\mtt{y} \mapsto 4$, $\mtt{s} \mapsto \text{''\texttt{o}''}$, $\mtt{t} \mapsto \text{''\texttt{k}''}$. The lower half of the diagram represents the application of the instance $\mathtt{bridge}^{g,\alpha}$ according to $g$. Note that the application condition of \texttt{bridge} (see  Figure \ref{fig:condruleschema}) is satisfied with respect to $g$ and $\alpha$.

In the remainder of this section, we make the above four steps precise. Consider a conditional rule schema $r = \tuple{L \gets K \to R,\, c}$. Given graphs $G$\/ in $\G(\mrm{Label})$ and $H$\/ in $\G(\L)$, a \emph{premorphism} $g\colon G \to H$\/ consists of two functions $g_V\colon V_G \to V_H$ and $g_E\colon E_G \to E_H$\/ that preserve sources and targets: $s_H \circ g_E = g_V \circ s_G$ and $t_H \circ g_E = g_V \circ t_G$. 

An \emph{assignment}\/ is a family of mappings $\alpha = (\alpha_X)_{X \in \{\mrm{I},\mrm{S},\mrm{A},\mrm{L}\}}$ where $\alpha_{\mrm{I}}\colon \mrm{IVariable} \to \Z$, $\alpha_{\mrm{S}}\colon \mrm{SVariable}\linebreak[3] \to \C^*$, $\alpha_{\mrm{A}}\colon \mrm{AVariable} \to \Z \cup \C^*$ and $\alpha_{\mrm{L}}\colon \mrm{LVariable} \to \L$. We sometimes omit the subscripts of these mappings as exactly one of them is applicable to a given variable.

Given a premorphism $g\colon L \to G$, an assignment $\alpha$ and a label $l = e\, m$ with $e \in \mrm{List}$ and $m \in \{\mtt{true},\mtt{false}\}$, the value $l^{g,\alpha} \in \L$ is the pair $\tuple{e^{g,\alpha},\,\mrm{true}}$ if $m = \mtt{true}$ and $\tuple{e^{g,\alpha},\,\mrm{false}}$ otherwise.

The value $e^{g,\alpha} \in (\Z \cup \C^*)^*$ is inductively defined. If $e = \mtt{empty}$, then $e^{g,\alpha}$ is the empty sequence. If $e$\/ has the form $d_1 \dots d_n$ ($n \geq 1$) with digits $d_1,\dots,d_n$, or has the form ''$c_1 \dots c_n$'' ($n \geq 0$) with characters $c_1,\dots,c_n$, then $e^{g,\alpha}$ is the unique integer in $\Z$ resp.\ character string in $\C^*$ represented by $e$. (Note that the empty character string and $\mtt{empty}^{g,\alpha}$ are different values.) If $e$\/ is a variable, then $e^{g,\alpha} = \alpha(e)$. Otherwise, $e^{\alpha}$ is obtained from the values of $e$'s components. If $e = -e_1$ with $e_1 \in \mrm{Integer}$, then $e^{g,\alpha}$ is the integer opposite to $e_1^{g,\alpha}$. If $e$\/ has the form $e_1 \oplus e_2$ with $\oplus \in \mrm{ArithOp}$ and $e_1,e_2 \in \mrm{Integer}$, then $e^{g,\alpha} = e_1^{g,\alpha} \oplus_{\Z} e_2^{g,\alpha}$ where $\oplus_{\Z}$ is the integer operation represented by $\oplus$.\footnote{The effect of dividing by zero is undefined, that is, left to the implementation.} If $e$\/ has the form $\mtt{indeg}(n)$ or $\mtt{outdeg}(n)$, with $n \in \mrm{Node}$, then $e^{g,\alpha}$ is the indegree resp.\ outdegree of the node $g_V(n)$ in $G$. Finally, if $e = e_1 \mtt{.} e_2$ with $e_1,e_2 \in \mathrm{String}$ or $e = e_1 \mtt{:} e_2$ with $e_1,e_2 \in \mathrm{List}$, then $e^{g,\alpha}$ is the concatenation of $e_1^{g,\alpha}$ and $e_2^{g,\alpha}$. 

The value $c^{g,\alpha} \in \B$ of the condition $c$ is also inductively defined. If $c$\/ has the form $\mtt{int}(e_1)$ with $e_1 \in \mathrm{List}$, then $c^{g,\alpha} = \mrm{true}$ if and only if $e_1^{g,\alpha} \in \Z$. Similarly, if  $c$\/ has the form $\mtt{string}(e_1)$ or $\mtt{atom}(e_1)$, then $c^{g,\alpha} = \mrm{true}$ if and only if $e_1^{g,\alpha} \in \C^*$ resp.\ $e_1^{g,\alpha} \in \Z \cup \C^*$. If $c$\/ has the form $e_1 \mathop{\mtt{=}} e_2$ or $e_1 \mathop{\mtt{!=}} e_2$ with $e_1,e_2 \in \mrm{List}$, then $c^{g,\alpha} = \mrm{true}$ if and only if $e_1^{g,\alpha} = e_2^{g,\alpha}$ resp.\ $e_1^{g,\alpha} \neq e_2^{g,\alpha}$. If $c$\/ has the form $e_1 \bowtie e_2$ with $\mathop{\bowtie} \in \mrm{RelOp}$ and $e_1,e_2$ in $\mrm{Integer}$, then $c^{g,\alpha} = \mrm{true}$ if and only if $e_1^{g,\alpha} \bowtie_{\Z} e_2^{g,\alpha}$ where $\bowtie_{\Z}$ is the integer relation represented by $\bowtie$. 

If $c$ has the form $\mathtt{edge}(m,n)$ with $m,n \in \mrm{Node}$, then $c^{g,\alpha} = \mrm{true}$ if and only if there is an edge in $G$ from $g_V(m)$ to $g_V(n)$. Similarly, if $c$ has the form $\mathtt{edge}(m,n,e)$ with $m,n \in \mrm{Node}$ and $e \in \mrm{List}$ , then $c^{g,\alpha} = \mrm{true}$ if and only if there is an edge from $g_V(m)$ to $g_V(n)$ with a label whose list component is $e^{g,\alpha}$.

If $c$ has the form $\mathop{\mtt{not}} c_1$ with $c_1 \in \mrm{Condition}$, then $c^{g,\alpha} = \mrm{true}$ if and only if $c_1^{g,\alpha} = \mrm{false}$. Finally, if $c$ has the form $c_1 \mathop{\mtt{and}} c_2$ with $c_1,c_2 \in \mrm{Condition}$, then $c^{g,\alpha} = \mrm{true}$ if and only if $c_1^{g,\alpha} = \mrm{true}= c_2^{g,\alpha}$, and  if $c$ has the form $c_1 \mathop{\mtt{or}} c_2$, then $c^{g,\alpha} = \mrm{true}$ if and only $c_1^{g,\alpha} = \mrm{true}$ or $c_2^{g,\alpha} = \mrm{true}$.

We call $r^{g,\alpha} = \tuple{L^{g,\alpha} \gets K \to R^{g,\alpha}}$ the \emph{instance}\/ of $r$ with respect to $g$ and $\alpha$, where  $L^{g,\alpha}$ and $R^{g,\alpha}$ are obtained from $L$ and $R$ by replacing each label $l$ with $l^{g,\alpha}$. Note that $r^{g,\alpha}$ is a graph transformation rule over $\L$, in the sense of Section \ref{sec:newgratra}. We can now define the application of conditional rule schemata to graphs in $\G(\L)$.

\begin{definition}[Rule-schema application]
\label{def:condruleschema_appl}
Given a conditional rule schema $r = \tuple{L \gets K \to R,\, c}$ and graphs $G,H$ in $\G(\L)$, we write $G \dder_{r,g} H$\/ (or just $G \dder_r H$) if there are a premorphism $g\colon L \to G$\/ and an assignment $\alpha$ such that 
\begin{enumerate}
\item[(1)] $g$ is a graph morphism $L^{g,\alpha} \to G$, 
\item[(2)] $c^{g,\alpha} = \mrm{true}$, and 
\item[(3)] $G \dder_{r^{g,\alpha},g} H$.
\end{enumerate}
\end{definition}
Here $G \dder_{r^{g,\alpha},g} H$\/ denotes the application of $r^{g,\alpha}$ with match $g$ to $G$, as defined in Section \ref{sec:newgratra}. Note that we use $\dder$ for the application of both rule schemata and rules, to avoid an inflation of symbols. Given a set $\R$ of conditional rule schemata, we write $G \dder_{\R} H$\/ if $G \dder_r H$\/ for some  conditional rule schema $r$\/ in $\R$.

The following proposition shows that given a rule schema $r$, a premorphism from the left-hand graph of $r$ to $G$\/ induces at most one instance of $r$\/ that can be applied with match $g$.
\begin{proposition}
Given a conditional rule schema $r = \tuple{L \gets K \to R,\, c}$ and a premorphism $g\colon L \to G$, there exists at most one assignment $\alpha$ such that $g$ is a graph morphism $L^{g,\alpha} \to G$.
\end{proposition}
The proof of this property relies on the fact that the left-hand graph $L$\/ contains only simple expressions.

%% file: Graphs/condruleschema_appl.graph
   \setlength{\unitlength}{0.5in}
   \graphnodesize{0.9}
   \grapharrowlength{0.2}
   \grapharrowwidth{0.6}
   \autodistance{1.0}
   \graphnodecolour{1}

\begin{tabular}{rcl}
\begin{graph}(3.8,1.8)(.2,1)
 \roundnode{n1}(.6,1.7)[\grey]
  \autonodetext{n1}{\texttt{a:x}}\autonodetext{n1}[s]{\tiny 1}
 \roundnode{n2}(2.1,1.7)
  \autonodetext{n2}{\texttt{n}}\autonodetext{n2}[s]{\tiny 2}
 \roundnode{n3}(3.6,1.7)
  \autonodetext{n3}{\texttt{y}}\autonodetext{n3}[s]{\tiny 3}
 \diredge{n1}{n2}\bowtext{n1}{n2}{.12}{\texttt{s}}
 \diredge{n2}{n3}\bowtext{n2}{n3}{.12}{\texttt{t}}
\end{graph}
&
\begin{graph}(1.4,1.8)(0,.07)
 \freetext(.7,.75){$\dder$}
\end{graph}
&
\begin{graph}(3.8,1.8)(.2,1)
 \roundnode{n1}(.6,1.7)
  \autonodetext{n1}{\texttt{a}}\autonodetext{n1}[s]{\tiny 1}
  \roundnode{n2}(2.1,1.7)
  \autonodetext{n2}{\texttt{x:n}}\autonodetext{n2}[s]{\tiny 2}
 \roundnode{n3}(3.6,1.7)[\grey]\autonodetext{n3}[s]{\tiny 3}
  \autonodetext{n3}{$\mtt{n \ast n}$}\autonodetext{n3}[s]{\tiny 3}
 \dirbow{n1}{n3}{.25}[\dashed]\bowtext{n1}{n3}{.31}{$\mathtt{s.t}$}
 \diredge{n1}{n2}\bowtext{n1}{n2}{.12}{\texttt{s}}
 \diredge{n2}{n3}\bowtext{n2}{n3}{.12}{\texttt{t}}
\end{graph}
\\[1ex]
\begin{graph}(3.8,.5)
  \freetext(1.9,.2)%
  {\begin{turn}{-90} \mbox{\large $\mapsto$} \end{turn}}
  \freetext(2.3,.2){\small $\alpha, g$}
\end{graph}
&
\begin{graph}(1.4,.5)
\end{graph}
&
\begin{graph}(3.8,.5)
  \freetext(1.9,.2)%
  {\begin{turn}{-90} \mbox{\large $\mapsto$} \end{turn}}
  \freetext(2.3,.2){\small $\alpha, g$}
\end{graph}
\\[2ex]
\begin{graph}(3.8,1.8)(.2,1)
 \roundnode{n1}(.6,1.7)[\grey]
  \autonodetext{n1}{\texttt{0:1:2}}[\grey]\autonodetext{n1}[s]{\tiny 1}
 \roundnode{n2}(2.1,1.7)
  \autonodetext{n2}{\texttt{3}}\autonodetext{n2}[s]{\tiny 2}
 \roundnode{n3}(3.6,1.7)
  \autonodetext{n3}{\texttt{4}}\autonodetext{n3}[s]{\tiny 3}
 \diredge{n1}{n2}\bowtext{n1}{n2}{.15}{''\texttt{o}''}
 \diredge{n2}{n3}\bowtext{n2}{n3}{.15}{''\texttt{k}''}
\end{graph}
&
\begin{graph}(1.4,1.8)
 \freetext(.7,.7){$\dder$}
\end{graph}
&
\begin{graph}(3.8,1.8)(.2,1)
 \roundnode{n1}(.6,1.7)
  \autonodetext{n1}{\texttt{0}}\autonodetext{n1}[s]{\tiny 1}
  \roundnode{n2}(2.1,1.7)
  \autonodetext{n2}{\texttt{1:2:3}}\autonodetext{n2}[s]{\tiny 2}
 \roundnode{n3}(3.6,1.7)[\grey]
  \autonodetext{n3}{\texttt{9}}\autonodetext{n3}[s]{\tiny 3}
 \dirbow{n1}{n3}{.25}[\dashed]\bowtext{n1}{n3}{.31}{''\texttt{ok}''}
 \diredge{n1}{n2}\bowtext{n1}{n2}{.15}{''\texttt{o}''}
 \diredge{n2}{n3}\bowtext{n2}{n3}{.15}{''\texttt{k}''}
\end{graph}
\\[1ex]
\begin{graph}(3.8,.5)
  \freetext(1.9,.2){$\downarrow$}
  \freetext(2.15,.2){\small $g$}
\end{graph}
&
\begin{graph}(1.4,.5)
\end{graph}
&
\begin{graph}(3.8,.5)
  \freetext(1.9,.2){$\downarrow$}
\end{graph}
\\[2ex]
\begin{graph}(3.8,2.8)(.2,0)
 \roundnode{n1}(.6,1.7)[\grey]
  \autonodetext{n1}{\texttt{0:1:2}}
 \roundnode{n2}(2.1,1.7)
  \autonodetext{n2}{\texttt{3}}
 \roundnode{n3}(3.6,1.7)
  \autonodetext{n3}{\texttt{4}}
 \roundnode{n4}(2.1,.5)
  \autonodetext{n4}{\texttt{2}}
 \diredge{n1}{n2}\bowtext{n1}{n2}{.15}{''\texttt{o}''}
 \diredge{n2}{n3}\bowtext{n2}{n3}{.15}{''\texttt{k}''}
 \diredge{n3}{n4}\bowtext{n3}{n4}{.15}{$\mtt{0}$}
 \diredge{n4}{n1}\bowtext{n4}{n1}{.15}{$\mtt{-1}$}
\end{graph}
&
\begin{graph}(1.4,2.8)(0,-.6)
 \freetext(.7,1.1){$\dder$}
\end{graph}
&
\begin{graph}(3.8,2.8)(.2,0)
 \roundnode{n1}(.6,1.7)
  \autonodetext{n1}{\texttt{0}}
  \roundnode{n2}(2.1,1.7)
  \autonodetext{n2}{\texttt{1:2:3}}
 \roundnode{n3}(3.6,1.7)[\grey]
  \autonodetext{n3}{\texttt{9}}
 \roundnode{n4}(2.1,.5)
  \autonodetext{n4}{\texttt{2}}
 \dirbow{n1}{n3}{.25}[\dashed]\bowtext{n1}{n3}{.31}{''\texttt{ok}''}
 \diredge{n1}{n2}\bowtext{n1}{n2}{.15}{''\texttt{o}''}
 \diredge{n2}{n3}\bowtext{n2}{n3}{.15}{''\texttt{k}''}
 \diredge{n3}{n4}\bowtext{n3}{n4}{.15}{$\mtt{0}$}
 \diredge{n4}{n1}\bowtext{n4}{n1}{.15}{$\mtt{-1}$}
\end{graph}
\end{tabular}

%% file: programs.tex
\section{Programs}
\label{sec:programs}

The syntax of graph programs is the same as in GP 1, except for the syntax of rule schemata and the new constructs \texttt{try\_then\_else} and \texttt{or}. Figure \ref{fig:program_syntax} shows the abstract syntax of GP 2 programs. As before, a program consists of a number of declarations of conditional rule schemata and macros, and exactly one declaration of a main command sequence. The identifiers of category RuleId occurring in a RuleSetCall refer to declarations of conditional rule schemata in category RuleDecl (see previous sections). 

\begin{figure}[htb]
\renewcommand{\arraystretch}{1.2}
\begin{center}
\begin{tabular}{lcl}
Prog & ::= & Decl \{Decl\} \\
Decl & ::= & RuleDecl $\mid$ MacroDecl $\mid$ MainDecl \\
MacroDecl & ::= & MacroId '=' ComSeq \\
MainDecl & ::= & \texttt{main} '=' ComSeq \\
ComSeq & ::= & Com \{';' Com\} \\
Com & ::= & RuleSetCall $\mid$ MacroCall \\
&& $\mid$ \texttt{if} ComSeq \texttt{then} ComSeq [\texttt{else} ComSeq] \\
&& $\mid$ \texttt{try} ComSeq \texttt{then} ComSeq [\texttt{else} ComSeq] \\
&& $\mid$ ComSeq '!' \\
&& $\mid$ ComSeq \texttt{or} ComSeq \\
&& $\mid$ \texttt{skip} $\mid$ \texttt{fail} \\
RuleSetCall & ::= & RuleId $\mid$ '\{' [RuleId \{',' RuleId\}] '\}' \\
MacroCall & ::= & MacroId 
\end{tabular}
\end{center}
\caption{Abstract syntax of programs}\label{fig:program_syntax}
\end{figure}

In the next section it is shown that the commands \texttt{or}, \texttt{skip} and \texttt{fail} can be expressed through the other commands. Hence the core of GP includes only the call of a set of conditional rule schemata (RuleSetCall), sequential composition (';'), the if-then-else statement, the try-then-else statement and as-long-as-possible iteration ('!'). 
Before formally defining the semantics of programs, we discuss some example programs to illustrate the use of the new features of GP 2.

\begin{example}[Checking connectedness]
\label{ex:connected}

A graph is \emph{connected}\/ if there is an undirected path between each two nodes, that is, a sequence of consecutive edges whose directions don't matter. The program in Figure \ref{fig:connected} checks whether an arbitrary input graph $G$ is connected and, depending on the result, executes either program $P$\/ or program $Q$ on $G$. 

Connectedness is checked by picking some node, marking it, and propagating node marks along edges as long as possible. Then an application of the rule schema \texttt{unmarked} tests whether any unmarked nodes are left. If this is the case, then the macro \texttt{disconnected} succeeds and program $Q$\/ is executed, otherwise \texttt{disconnected} fails and program $P$\/ is executed.

It is important to note that $P$\/ or $Q$ is executed \emph{on the input graph}\/ whereas the graph resulting from the test is discarded. The precise semantics of the branching command is given in Section \ref{sec:sos}.
\qed

\begin{figure}[htb]
 \begin{center}
  \input{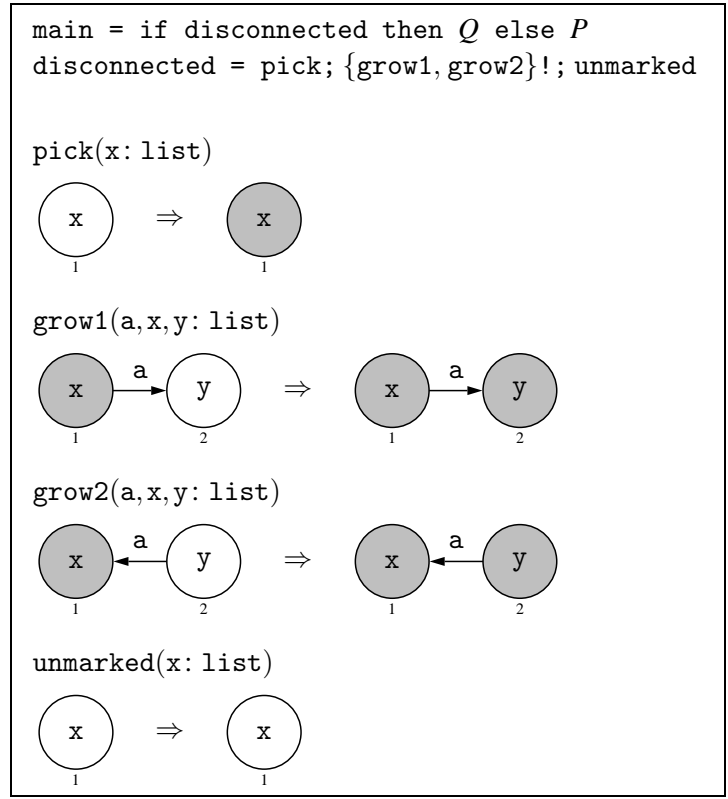}
 \end{center}
\caption{A program for checking connectedness}
\label{fig:connected}
\end{figure}

\end{example}

\begin{example}[Recognising acyclic graphs]
\label{ex:acyclic}

A graph is \emph{acyclic}\/ if it does not contain a directed cycle. The program in Figure \ref{fig:acyclic} checks whether an unmarked input graph $G$ is acyclic and, depending on the result, executes either program $P$\/ or program $Q$ on $G$. 

The absence of cycles is checked by deleting, as long as possible, edges whose source nodes have no incoming edges, and testing subsequently whether any edges remain. This method relies on the following invariant of the rule schema \texttt{delete}: for every step $G \dder_{\mtt{delete}} H$, $G$ is acyclic if and only if $H$\/ is acyclic. Moreover, a graph to which \texttt{delete} is not applicable is acyclic if and only if it does not contain edges. Note that the condition of \texttt{delete} uses the new indegree function.
\qed

\begin{figure}[htb]
 \begin{center}
  \input{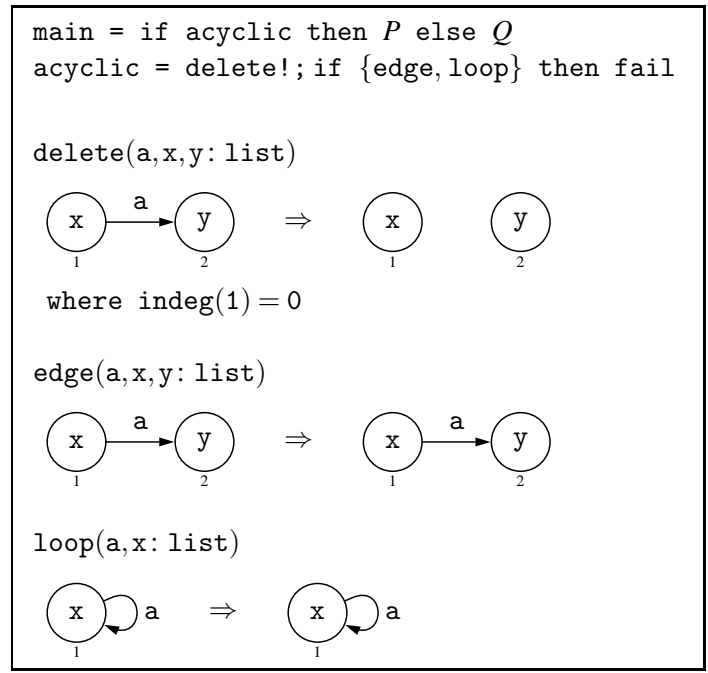}
 \end{center}
\caption{A program for recognising acyclic graphs}
\label{fig:acyclic}
\end{figure}

\end{example}

\begin{example}[Recognising series-parallel graphs]
\label{ex:series-parallel}

\emph{Series-parallel}\/ graphs are inductively defined as follows. Every graph $G$ consisting of two nodes connected by an edge is series-parallel, where the edge's source and target are the source and target of $G$. Given series-parallel graphs $G$ and $H$, the graphs obtained from the disjoint union $G + H$\/ by the following two operations are also series-parallel. Serial composition: merge the target of $G$ with the source of $H$; the source of $G$ becomes the new source and the target of $H$\/ becomes the new target. Parallel composition: merge the source of $G$ with the source of $H$, and the target of $G$ with the target of $H$; sources and targets are preserved.

It is known \cite{Bang_Jensen-Gutin09a,Duffin65a} that a graph is series-parallel if and only if it reduces to a graph consisting of two nodes connected by an edge (a \emph{base graph}\/) by repeated application of the following operations: (a) Given a node with one incoming edge $i$ and one outgoing edge $o$ such that $s(i) \neq t(o)$, replace $i$, $o$ and the node by an edge from $s(i)$ to $t(o)$. (b) Replace a pair of parallel edges by an edge from their source to their target. 

\begin{figure}[htb]
 \begin{center}
  \input{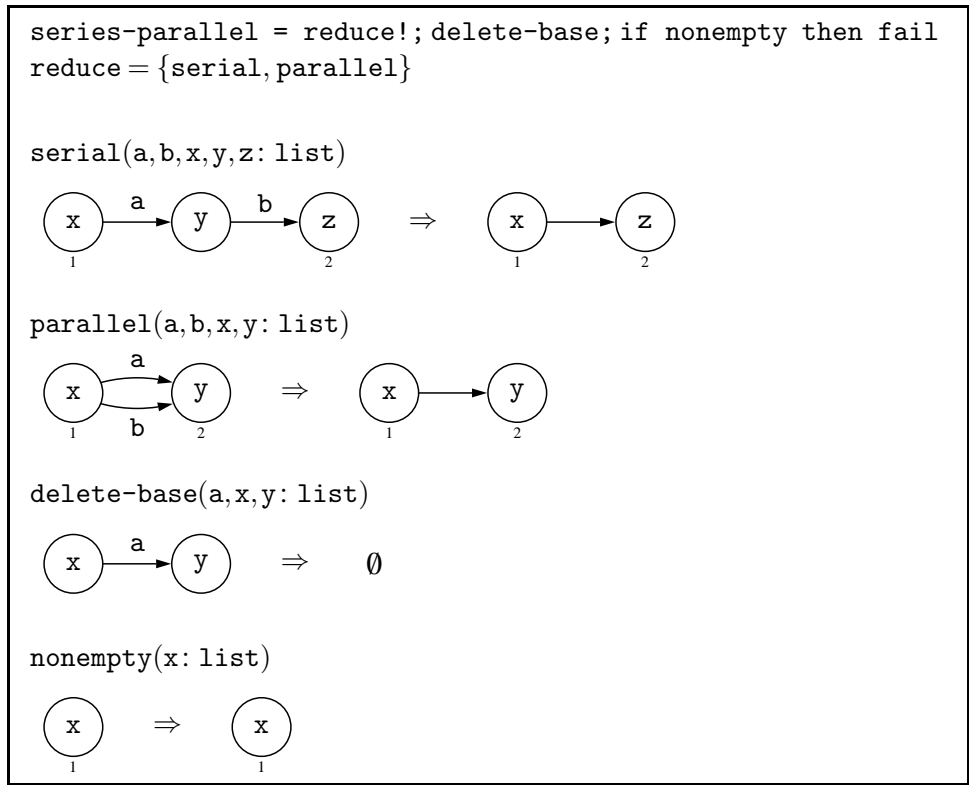}
 \end{center}
\caption{A macro for recognising series-parallel graphs}
\label{fig:series-parallel}
\end{figure}

Figure \ref{fig:series-parallel} shows a macro which reduces every unmarked series-parallel graph to the empty graph, and fails on every other unmarked graph. The subprogram \texttt{reduce!}\ applies as long as possible the operations (a) and (b) to the input graph $G$, then the rule schema \texttt{delete-base} checks if the result is a base graph whose nodes are not incident to other edges. (The latter is ensured by the dangling condition.) If \texttt{delete-base} is not applicable, then the input graph was not reduced to a base graph. In this case the input graph is not series-parallel because every execution of \texttt{reduce!}\ yields the same graph. (This is because the critical pairs of the rule schemata \texttt{serial} and \texttt{parallel} are strongly joinable \cite{Plump05a}.)

Finally, after \texttt{delete-base} has been applied, the rule schema \texttt{nonempty} checks whether the graph resulting from \texttt{reduce!} contains nodes other than those of the base graph. The input graph is series-parallel if and only if this is not the case. 
\qed
\end{example}

\begin{example}[Computing Euler cycles]
\label{ex:euler-cycles}

An \emph{Euler cycle}\/ is a directed cycle of distinct edges that contains all edges and nodes of a graph. A graph is \emph{eulerian}\/ if it contains an Euler cycle. It is known that a graph is eulerian if and only if it is connected and each node has the same indegree as outdegree \cite{Bang_Jensen-Gutin09a}. Based on this characterisation, the macro \texttt{eulerian} of Figure \ref{fig:eulerian} checks whether an unmarked graph is eulerian or not. It does this by using the macro \texttt{disconnected} of Figure \ref{fig:connected} and the new indegree and outdegree functions.

\begin{figure}[htb]
 \begin{center}
  \input{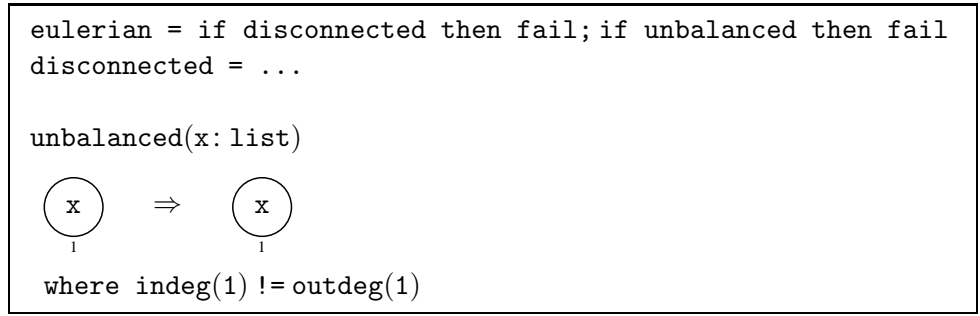}
 \end{center}
\caption{A macro for recognising eulerian graphs}
\label{fig:eulerian}
\end{figure}

Given an unmarked eulerian input graph with atomic labels, the program in Figure \ref{fig:euler-cycle} computes an Euler cycle and numbers its edges. 
\begin{figure}[htbp]
 \begin{center}
  \input{Programs/euler-cycle.prog}
 \end{center}
\caption{A program for computing an Euler cycle}
\label{fig:euler-cycle}
\end{figure}
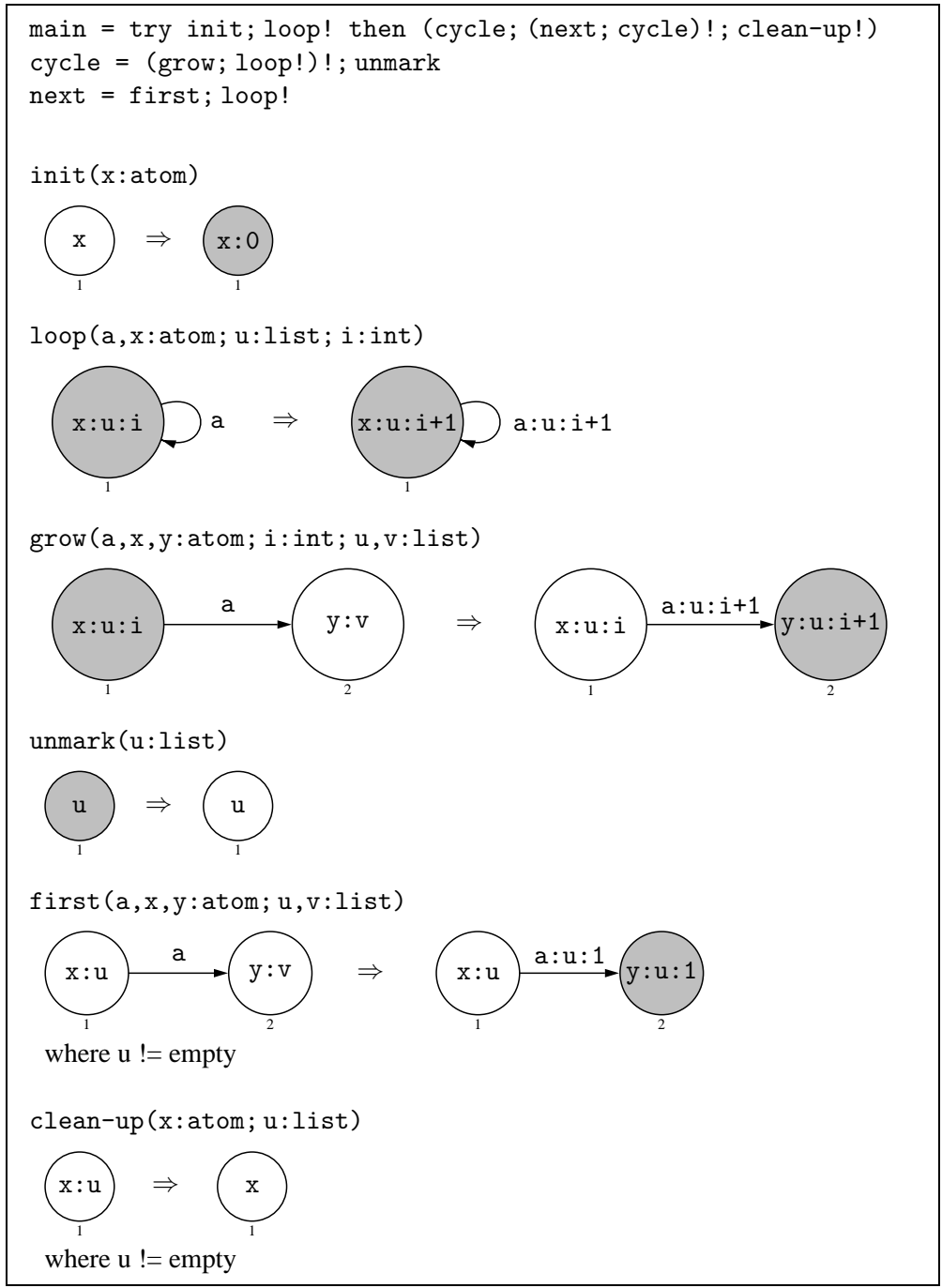
An execution of this program is shown in Figure \ref{fig:euler-execution}. In the resulting graph, the computed Euler cycle is given by the edges with the labels \ttt{1:1}, \ttt{1:1:1}, \ttt{1:1:2}, \ttt{1:1:3}, \ttt{1:2}, \ttt{1:3} and \ttt{1:4}. The graph in the middle of Figure \ref{fig:euler-execution} is an intermediate result representing the point in time when the macro \ttt{cycle} has been executed for the first time.
\begin{figure}[htb]
 \begin{center}
  \input{Graphs/euler-execution.graph}
 \end{center}
\caption{An execution of the program of Figure \ref{fig:euler-cycle}}
\label{fig:euler-execution}
\end{figure}
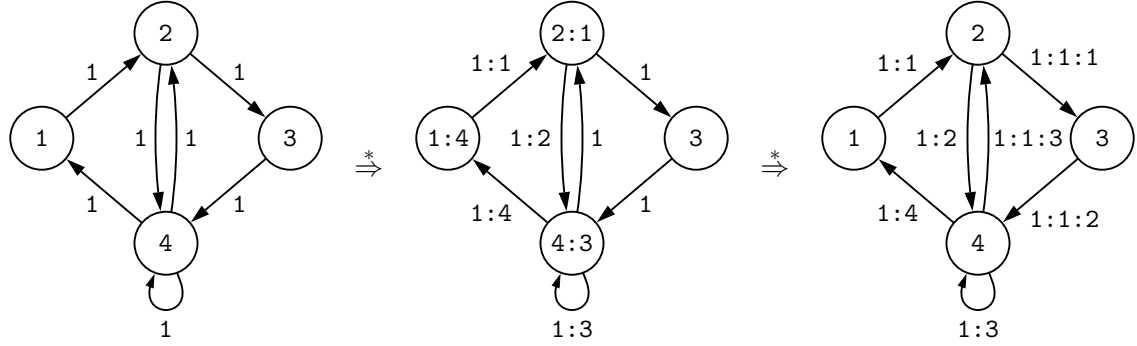

The program uses the command \ttt{try\_then} to check if the input graph is nonempty. If the input graph is empty, then the empty sequence of edges is an Euler cycle and hence the program returns the empty graph. If the input graph is nonempty, the rule schema \ttt{init} picks some node, adds \ttt{0} to its label, and marks the node. Then the rule schema \ttt{loop} numbers all loops with atomic labels that are incident to the node. Next the rule schema \ttt{cycle} numbers a proper (that is, non-loop) cycle starting at this node, by repeatedly applying the rule schema \ttt{grow}. Also, at each visited node, \ttt{loop} is applied as long as possible to number all incident loops. 

When the first proper cycle has been numbered, the subprogram \ttt{(next;{\csp}cycle)!} repeatedly computes a new cycle starting at a node that has already been visited. This cycle is inserted into the current cycle by numbering the new edges with lists that add one position to the list of the edge preceding the new edges. Finally, when all edges of the graph have been numbered, the rule schema \ttt{clean-up} removes all auxiliary information in node labels. 
\qed
\end{example}

%% file: Programs/euler-cycle.prog
   \graphnodesize{1}
   \grapharrowlength{0.2}
   \grapharrowwidth{0.6}
   \autodistance{1.0}
   \graphnodecolour{1}

\fbox{%
\begin{tabular}{l}
\ttt{main = try~init;{\csp}loop!~then~(cycle;{\csp}(next;{\csp}cycle)!;{\csp}clean-up!)}\\
\ttt{cycle = (grow;{\csp}loop!)!;{\csp}unmark}\\
\ttt{next = first;{\csp}loop!}
\\\\[1ex]
\ttt{init(x:atom)}\\[1ex]
\begin{tabular}{rcl}
\begin{graph}(1,1)(0,-.1)
\roundnode{A}(.5,.4)
\autonodetext{A}{$\x$}
\autonodetext{A}[s]{\tiny 1}
\end{graph}
&
\begin{graph}(.4,1)
\freetext(.2,0.5){$\dder$}
\end{graph}
&
\begin{graph}(1,1)
\roundnode{A}(.5,.5)[\grey]
\autonodetext{A}{\ttt{x:0}}
\autonodetext{A}[s]{\tiny 1}
\end{graph}
\end{tabular}
\\\\
\ttt{loop(a,x:atom;{\csp}u:list;{\csp}i:int)}\\[1ex]
\begin{tabular}{rcl}
\begin{graph}(2.6,1.6)(-.4,0)
\roundnode{A}(.5,.8)[\graphnodesize{1.6}\grey]
\autonodetext{A}{\ttt{x:u:i}}
\autonodetext{A}[s]{\tiny 1}
\dirloopedge{A}(.75,.25)(.75,-.25)
\freetext(2.05,.8){\ttt{a}}
\end{graph}
&
\begin{graph}(.8,1.6)
\freetext(.4,.8){$\dder$}
\end{graph}
&
\begin{graph}(3.9,1.6)(-.4,0)
\roundnode{A}(.5,.8)[\graphnodesize{1.6}\grey]
\autonodetext{A}{\ttt{x:u:i+1}}
\autonodetext{A}[s]{\tiny 1}
\dirloopedge{A}(.75,.25)(.75,-.25)
\freetext(2.7,.8){\ttt{a:u:i+1}}
\end{graph}
\end{tabular}
\\\\
\ttt{grow(a,x,y:atom;{\csp}i:int;{\csp}u,v:list)}\\[1ex]
\begin{tabular}{rcl}
\begin{graph}(5,1.6)(-.4,0)
\roundnode{A}(.5,.8)[\graphnodesize{1.6}\grey]
\autonodetext{A}{\ttt{x:u:i}}
\roundnode{B}(3.9,.8)[\graphnodesize{1.6}]
\autonodetext{B}{\ttt{y:v}}
\autonodetext{A}[s]{\tiny 1}
\autonodetext{B}[s]{\tiny 2} 
\diredge{A}{B}
\bowtext{A}{B}{.08}{\ttt{a}}
\end{graph}
&
\begin{graph}(1.2,1.6)
\freetext(.6,.8){$\dder$}
\end{graph}
&
\begin{graph}(5,1.6)(-.2,0)
\roundnode{A}(.5,.8)[\graphnodesize{1.6}]
\autonodetext{A}{\ttt{x:u:i}}
\roundnode{B}(3.9,.8)[\graphnodesize{1.6}\grey]
\autonodetext{B}{\ttt{y:u:i+1}}
\autonodetext{A}[s]{\tiny 1}
\autonodetext{B}[s]{\tiny 2} 
\diredge{A}{B}
\bowtext{A}{B}{.08}{\ttt{a:u:i+1}}
\end{graph}
\end{tabular}
\\\\
\ttt{unmark(u:list)}\\[1ex]
\begin{tabular}{rcl}
\begin{graph}(1,1)(.3,-.1)
\roundnode{A}(.8,.4)[\grey]
\autonodetext{A}{\ttt{u}}
\autonodetext{A}[s]{\tiny 1}
\end{graph}
&
\begin{graph}(.4,1)
\freetext(.2,0.5){$\dder$}
\end{graph}
&
\begin{graph}(1,1)(0,-.1)
\roundnode{A}(.5,.4)
\autonodetext{A}{\ttt{u}}
\autonodetext{A}[s]{\tiny 1}
\end{graph}
\end{tabular}
\\\\
\ttt{first(a,x,y:atom;{\csp}u,v:list)}\\[1ex]
\begin{tabular}{rcl}
\begin{graph}(3.8,1.2)(.9,0)
\roundnode{A}(1.5,.6)[\graphnodesize{1.2}]
\autonodetext{A}{\ttt{x:u}}
\roundnode{B}(4.1,.6)[\graphnodesize{1.2}]
\autonodetext{B}{\ttt{y:v}}
\autonodetext{A}[s]{\tiny 1}
\autonodetext{B}[s]{\tiny 2} 
\diredge{A}{B}
\bowtext{A}{B}{.1}{\ttt{a}}
\end{graph}
&
\begin{graph}(.8,1.2)
\freetext(0.4,.6){$\dder$}
\end{graph}
&
\begin{graph}(4,1.2)(.8,0)
\roundnode{A}(1.5,.6)[\graphnodesize{1.2}]
\autonodetext{A}{\ttt{x:u}}
\roundnode{B}(4.1,.6)[\graphnodesize{1.2}\grey]
\autonodetext{B}{\ttt{y:u:1}}
\autonodetext{A}[s]{\tiny 1}
\autonodetext{B}[s]{\tiny 2} 
\diredge{A}{B}
\bowtext{A}{B}{.1}{\ttt{a:u:1}}
\end{graph}\\[1ex]
\multicolumn{3}{l}{\text{where u !=\/ empty}}
\end{tabular}
\\\\
\ttt{clean-up(x:atom;{\csp}u:list)}\\[1ex]
\begin{tabular}{rcl}
\begin{graph}(1,1)
\roundnode{A}(.5,.5)
\autonodetext{A}{\ttt{x:u}}
\autonodetext{A}[s]{\tiny 1}
\end{graph}
&
\begin{graph}(.6,1)
\freetext(.3,0.5){$\dder$}
\end{graph}
&
\begin{graph}(1,1)
\roundnode{A}(.5,.5)
\autonodetext{A}{\ttt{x}}
\autonodetext{A}[s]{\tiny 1}
\end{graph}\\[1ex]
\multicolumn{3}{l}{\text{where u !=\/ empty}}
\end{tabular}
\end{tabular}
}

%% file: Graphs/euler-execution.graph
   \setlength{\unitlength}{0.5in}
   \graphnodesize{0.68}
   \grapharrowlength{0.2}
   \grapharrowwidth{0.6}
   \autodistance{1.0}
   \graphnodecolour{1}
   \opaquetextfalse

\begin{tabular}{c}
\begin{graph}(3.4,3.7)(0,-.7)
 \roundnode{n1}(.4,1.5)\autonodetext{n1}{\small\ttt{1}}
 \roundnode{n2}(1.7,2.6)\autonodetext{n2}{\small\ttt{2}}
 \roundnode{n3}(3,1.5)\autonodetext{n3}{\small\ttt{3}}
 \roundnode{n4}(1.7,.4)\autonodetext{n4}{\small\ttt{4}}
 \dirloopedge{n4}(.15,-.4)(-.15,-.4)[\grapharrowlength{0.15}]
 \freetext(1.7,-.5){\small\ttt{1}}
 \diredge{n1}{n2}\bowtext{n1}{n2}{.11}{\small\ttt{1}}
 \diredge{n2}{n3}\bowtext{n2}{n3}{.11}{\small\ttt{1}}
 \diredge{n3}{n4}\bowtext{n3}{n4}{.11}{\small\ttt{1}}
 \diredge{n4}{n1}\bowtext{n4}{n1}{.11}{\small\ttt{1}}
 \dirbow{n2}{n4}{-.05}\bowtext{n2}{n4}{-.12}{\small\ttt{1}}
 \dirbow{n4}{n2}{-.05}\bowtext{n4}{n2}{-.12}{\small\ttt{1}}
\end{graph}
\begin{graph}(.7,3.7)
 \freetext(.35,1.95){$\DSdder^*$}
\end{graph}
\begin{graph}(3.4,3.7)(0,-.7)
 \roundnode{n1}(.4,1.5)\autonodetext{n1}{\small\ttt{1:4}}
 \roundnode{n2}(1.7,2.6)\autonodetext{n2}{\small\ttt{2:1}}
 \roundnode{n3}(3,1.5)\autonodetext{n3}{\small\ttt{3}}
 \roundnode{n4}(1.7,.4)\autonodetext{n4}{\small\ttt{4:3}}
 \dirloopedge{n4}(.15,-.4)(-.15,-.4)[\grapharrowlength{0.15}]
 \freetext(1.7,-.5){\small\ttt{1:3}}
 \diredge{n1}{n2}\bowtext{n1}{n2}{.18}{\small\ttt{1:1}}
 \diredge{n2}{n3}\bowtext{n2}{n3}{.11}{\small\ttt{1}}
 \diredge{n3}{n4}\bowtext{n3}{n4}{.11}{\small\ttt{1}}
 \diredge{n4}{n1}\bowtext{n4}{n1}{.18}{\small\ttt{1:4}}
 \dirbow{n2}{n4}{-.05}\bowtext{n2}{n4}{-.2}{\small\ttt{1:2}}
 \dirbow{n4}{n2}{-.05}\bowtext{n4}{n2}{-.12}{\small\ttt{1}}
\end{graph}
\begin{graph}(.7,3.7)
 \freetext(.35,1.95){$\DSdder^*$}
\end{graph}
\begin{graph}(3.4,3.7)(0,-.7)
 \roundnode{n1}(.4,1.5)\autonodetext{n1}{\small\ttt{1}}
 \roundnode{n2}(1.7,2.6)\autonodetext{n2}{\small\ttt{2}}
 \roundnode{n3}(3,1.5)\autonodetext{n3}{\small\ttt{3}}
 \roundnode{n4}(1.7,.4)\autonodetext{n4}{\small\ttt{4}}
 \dirloopedge{n4}(.15,-.4)(-.15,-.4)[\grapharrowlength{0.15}]
 \freetext(1.7,-.5){\small\ttt{1:3}}
 \diredge{n1}{n2}\bowtext{n1}{n2}{.18}{\small\ttt{1:1}}
 \diredge{n2}{n3}\bowtext{n2}{n3}{.23}{\small\ttt{1:1:1}}
 \diredge{n3}{n4}\bowtext{n3}{n4}{.23}{\small\ttt{1:1:2}}
 \diredge{n4}{n1}\bowtext{n4}{n1}{.18}{\small\ttt{1:4}}
 \dirbow{n2}{n4}{-.05}\bowtext{n2}{n4}{-.2}{\small\ttt{1:2}}
 \dirbow{n4}{n2}{-.05}\bowtext{n4}{n2}{-.235}{\small\ttt{1:1:3}}
\end{graph}
\end{tabular}

%% file: sos.tex
\section{Operational Semantics}
\label{sec:sos}

This section presents a formal semantics for GP 2 in the style of Plotkin's structural operational semantics \cite{Plotkin04a}. As usual for this approach, inference rules inductively define a small-step transition relation $\to$ on \emph{configurations}. In our setting, a configuration is either a command sequence together with a graph, just a graph or the special element fail:
\[ \to \;\; \subseteq \; (\text{ComSeq} \times \G(\L)) \times 
      ((\text{ComSeq} \times \G(\L)) \cup \G(\L) \cup \{\fail\}). \]
Configurations in $\text{ComSeq} \times \G(\L)$, given by a rest program and a state in the form of a graph, represent states of unfinished computations while graphs in $\G(\L)$ are proper results. In addition, the element fail represents a failure state. A configuration $\gamma$ is said to be \emph{terminal}\/ if there is no configuration $\delta$ such that $\gamma \to \delta$.

Figure \ref{fig:core_sos_rules} in the Appendix shows the inference rules for the core commands of GP 2. Each rule consists of a premise and a conclusion separated by a horizontal bar. Both parts contain meta-variables for command sequences and graphs, where $R$ stands for a call in category RuleSetCall, $C,P,P',Q$ stand for command sequences in category ComSeq, and $G,H$\/ stand for graphs in $\G(\L)$. Meta-variables are considered to be universally quantified. For example, the rule $\mathrm{[call_1]}$ reads: ``For all $R$ in $\mathrm{RuleSetCall}$ and all $G,H$ in $\G(\L)$, $G \dder_R H$\/ implies $\tuple{R,\, G} \to H$.''
The transitive and reflexive-transitive closures of $\to$ are written $\to^+$ and $\to^*$, respectively. The notation $G \not\dder_R$ expresses that for graph $G$ in $\G(\L)$ there is no graph $H$\/ such that $G \dder_R H$. 

The if-then-else command has been designed to ``hide'' destructive tests. In Example \ref{ex:connected}, for instance, the test of the if-then-else command produces a graph with marked nodes. By the inference rules $\mathrm{[if_1]}$ and $\mathrm{[if_2]}$, this graph is discarded and program $P$ or $Q$ is executed on the input graph. In contrast, a program $\tryte{C}{P}{Q}$ passes any graph resulting from its test to $P$. If test $C$\/ fails, however, $Q$ is executed on the input graph.

The semantics of the if-then-else command and the as-long-as-possible loop in GP 1 have been modified to allow an efficient implementation. Previously, the conditions of branching commands and the bodies of loops were tested, in the worst case, by trying all possible executions starting from the current graph. This made branching and loop commands impractical for complex tests or large input graphs. In GP 2, the semantics of $\ifte{C}{P}{Q}$, $\tryte{C}{P}{Q}$, and $B\mtt{!}$ do not enforce backtracking when $C$\/ or $B$\/ fails. Instead, control is passed to program $Q$\/ or the loop is terminated, respectively. Note that this change increases the nondeterminism of evaluation in cases where $C$\/ or $B$\/ can both succeed and fail on the input graph.

The inference rules for the remaining GP commands are given in Figure \ref{fig:derived_sos_rules} of the Appendix. These commands are referred to as \emph{derived}\/ commands because they can be defined by the core commands, as shown below. 

The meaning of GP 2 programs is summarised by the semantic function $\Sem{\_}$ which assigns to each program $P$\/ the function $\Sem{P}$ mapping an input graph $G$ to the set of all possible results of executing $P$\/ on $G$. The application of $\Sem{P}$ to $G$ is written $\Sem{P}G$. The result set may contain, besides proper results in the form of graphs, the special values fail and $\bot$. The value fail indicates a failed program run while $\bot$ indicates a run that does not terminate or gets stuck. Program $P$ \emph{can diverge from}\/ $G$ if there is an infinite sequence $\tuple{P,\,G} \to \tuple{P_1,\,G_1} \to \tuple{P_2,\,G_2} \to \dots$ Also, $P$ \emph{can get stuck from}\/ $G$ if there is a terminal configuration $\tuple{Q,\,H}$ such that $\tuple{P,\,G} \to^* \tuple{Q,\,H}$.

\begin{definition}[Semantic function]
\label{def:semantic_function}
The \emph{semantic function} $\Sem{\_}\colon \mathrm{ComSeq} \to (\G(\L) \to 2^{\G(\L)\cup\{\fail,\bot\}})$ is defined by
   \[\Sem{P}G\, =\, \{X \in (\G(\L) \cup \{\fail\}) \mid \tuple{P,\,G}\DSto^+ X\} 
     \cup \{\bot \mid \text{$P$ can diverge or get stuck from $G$}\}. \]
\end{definition} 

In the current implementation of GP, reaching the failure state triggers backtracking which then attempts to find a proper result \cite{Manning-Plump08b}. However, backtracking can be switched off by the user.

A program can get stuck in two situations: (1) it contains a command $\ifte{C}{P}{Q}$ or $\tryte{C}{P}{Q}$ such that $C$\/ can diverge from some graph $G$ and can neither produce a proper result from $G$ nor fail from $G$, or (2) it contains a loop $B!$ whose body $B$ possesses the said property of $C$. The evaluation of such commands gets stuck because none of the inference rules for if-then-else, try-then-else or iteration is applicable. 

The semantic function of Definition \ref{def:semantic_function} suggests a straightforward notion of program  equivalence. 

\begin{definition}[Semantic equivalence]
\label{def:semantic_equivalence}
Two programs $P$\/ and $Q$\/ are \emph{semantically equivalent}, denoted by $P\equiv Q$, if $\Sem{P} = \Sem{Q}$.
\end{definition} 

For example, it is easy to see that the following equivalences between derived commands and core commands hold (where $\emptyset$ is the empty graph):
\begin{itemize}
\item $\mtt{skip} \equiv \mtt{null}$, where \texttt{null} is the rule schema $\emptyset \dder \emptyset$;
\item $\mtt{fail} \equiv \mtt{\{\}}$, where $\{\}$ is the empty set of rule schemata;
\item $\ift{C}{P}\, \equiv\, \ifte{C}{P}{\mtt{null}}$, for all programs $C$\/ and $P$;
\item $\tryt{C}{P}\, \equiv\, \tryte{C}{P}{\mtt{null}}$, for all programs $C$\/ and $P$.
 \end{itemize}

\enlargethispage{\baselineskip}

Less obvious is the following equivalence, showing that \texttt{or} is a derived command:
\[ P\mspace{.5mu} \mathop{\mtt{or}} Q\, \equiv\, \ifte{\mtt{remove!;\; \{create,\, null \};\; zero}}{P}{Q}, \]
 for all programs $P$\/ and $Q$. Here \texttt{remove} is a set of three rule schemata that delete arbitrary edges, loops and isolated nodes, \texttt{create} is the rule schema
\[ \mbox{\large $\emptyset$} \dder \nodegraph{0} \]
and \texttt{zero} is the rule schema
\[ \nodegraph{0} \dder \nodegraph{0}. \]

The following non-equivalence may be surprising, too:
\[ \tryte{C}{P}{Q}\, \not\equiv\, \ifte{C}{C;P}{Q}. \]
To witness, choose $C = \mtt{skip}\, \mathop{\mtt{or}}\,  \mtt{fail}$,  $P = \mtt{skip}$ and $Q = \mtt{skip}$. Then the \texttt{try}-program is equivalent to \texttt{skip} and hence cannot fail, but the \texttt{if}-program can fail.

%% file: conclusion.tex
\section{Conclusion}
\label{sec:conclusion}

GP allows high-level problem solving in the domain of graphs, by supporting rule-based programming and freeing programmers from dealing with low-level data structures for graphs. The language has a simple syntax and semantics, to facilitate both understanding by programmers and formal reasoning on programs. 

The revised language GP 2 has an improved type system, including list variables and subtypes, a new concept of marking nodes and edges graphically, new built-in functions for accessing the indegree and the outdegree of nodes, a more powerful \texttt{edge} predicate for conditions, new commands \texttt{try}-\texttt{then}-\texttt{else} and \texttt{or}, and a simplified semantics of branching and looping to enable an efficient implementation.

Topics for future work include the implementation of GP 2, tool support for Hoare-style program verification \cite{Poskitt-Plump12a}, and static analyses for properties such as termination and confluence.

\paragraph{\normalfont\textbf{Acknowledgements.}}
Parts of this paper were written while visiting Annegret Habel in Oldenburg and Berthold Hoffmann in Bremen in the autumn of 2011. I am grateful for their hospitality. Thanks go also to the Plasma research group in York for helpful comments, especially to Colin Runciman for proposing the concept of shaded nodes.

%% file: appendix.tex
\section*{Appendix: Semantic Inference Rules}
\label{sec:sos-rules}

\vspace{1ex}
\begin{figure}[htp]
\begin{center}
\begin{tabular}{lcl}
$\mathrm{[call_1]}$ $\frac{\displaystyle G \dder_R H}{\displaystyle\tuple{R,\,G} \to H}$ 
& \hspace{.5em} &
$\mathrm{[call_2]}$ $\frac{\displaystyle G \not\dder_R}{\displaystyle\tuple{R,\,G} \to \failrm}$
\\\\
$\mathrm{[seq_1]}$ $\frac{\displaystyle \tuple{P,\, G} \to \tuple{P',\, H}}{\displaystyle \tuple{P;Q,\, G} \to \tuple{P';Q,\, H}}$ 
&&
$\mathrm{[seq_2]}$ $\frac{\displaystyle \tuple{P,\, G} \to H}{\displaystyle \tuple{P;Q,\, G}\to \tuple{Q,\, H}}$
\\\\
$\mathrm{[seq_3]}$ $\frac{\displaystyle \tuple{P,\, G} \to \failrm}{\displaystyle \tuple{P;Q,\, G}\to \failrm}$
\\\\
$\mathrm{[if_1]}$ $\frac{\displaystyle \tuple{C,\, G} \to^+ H}{\displaystyle \tuple{\ifte{C}{P}{Q},\, G}\to \tuple{P,\, G}}$
&&
$\mathrm{[if_2]}$ $\frac{\displaystyle \tuple{C,\, G} \to^+ \failrm}{\displaystyle \tuple{\ifte{C}{P}{Q},\, G} \to \tuple{Q,\, G}}$
\\\\
$\mathrm{[try_1]}$ $\frac{\displaystyle \tuple{C,\, G} \to^+ H}{\displaystyle \tuple{\tryte{C}{P}{Q},\, G}\to \tuple{P,\, H}}$
&&
$\mathrm{[try_2]}$ $\frac{\displaystyle \tuple{C,\, G} \to^+ \failrm}{\displaystyle \tuple{\tryte{C}{P}{Q},\, G} \to \tuple{Q,\, G}}$
\\\\
$\mathrm{[alap_1]}$ $\frac{\displaystyle \tuple{P,\, G} \to^+ H}{\displaystyle \tuple{P!,\, G} \to \tuple{P!,\, H}}$
&&
$\mathrm{[alap_2]}$ $\frac{\displaystyle \tuple{P,\, G} \to^+ \failrm}{\displaystyle \tuple{P!,\, G} \to G}$
\end{tabular} 
\caption{Inference rules for core commands}\label{fig:core_sos_rules}
\end{center}

\vspace{3ex}
\begin{center}
\begin{tabular}{lcl}
$\mrm{[or_1]}$ $\tuple{P\mspace{.5mu} \mathop{\mtt{or}}\, Q,\, G} \to \tuple{P,\, G}$
& \hspace{.5em} &
$\mrm{[or_2]}$ $\tuple{P\mspace{.5mu} \mathop{\mtt{or}}\, Q,\, G} \to \tuple{Q,\, G}$
\\\\
$\mrm{[skip]}$ $\tuple{\skiptt,\, G} \to G$
&&
$\mrm{[fail]}$ $\tuple{\failtt,\, G} \to \failrm$
\\\\
$\mrm{[if_3]}$ $\frac{\displaystyle \tuple{C,\, G} \to^+ H}{\displaystyle \tuple{\ift{C}{P},\, G} \to \tuple{P,\, G}}$
&&
$\mathrm{[if_4]}$ $\frac{\displaystyle \tuple{C,\, G} \to^+ \failrm}{\displaystyle \tuple{\ift{C}{P},\, G} \to G}$
\\\\
$\mrm{[try_3]}$ $\frac{\displaystyle \tuple{C,\, G} \to^+ H}{\displaystyle \tuple{\tryt{C}{P},\, G} \to \tuple{P,\, H}}$
&&
$\mathrm{[try_4]}$ $\frac{\displaystyle \tuple{C,\, G} \to^+ \failrm}{\displaystyle \tuple{\tryt{C}{P},\, G} \to G}$
\end{tabular} 
\caption{Inference rules for derived commands}\label{fig:derived_sos_rules}
\end{center}
\end{figure}

%% file: article.bbl
\begin{thebibliography}{1}
\providecommand{\bibitemdeclare}[2]{}
\providecommand{\surnamestart}{}
\providecommand{\surnameend}{}
\providecommand{\urlprefix}{Available at }
\providecommand{\url}[1]{\texttt{#1}}
\providecommand{\href}[2]{\texttt{#2}}
\providecommand{\urlalt}[2]{\href{#1}{#2}}
\providecommand{\doi}[1]{doi:\urlalt{http://dx.doi.org/#1}{#1}}
\providecommand{\bibinfo}[2]{#2}

\bibitemdeclare{book}{Bang_Jensen-Gutin09a}
\bibitem{Bang_Jensen-Gutin09a}
\bibinfo{author}{J{\o}rgen \surnamestart Bang-Jensen\surnameend} \&
  \bibinfo{author}{Gregory \surnamestart Gutin\surnameend}
  (\bibinfo{year}{2009}): \emph{\bibinfo{title}{Digraphs: Theory, Algorithms
  and Applications}}, \bibinfo{edition}{second} edition.
\newblock \bibinfo{publisher}{Springer-Verlag}.

\bibitemdeclare{article}{Duffin65a}
\bibitem{Duffin65a}
\bibinfo{author}{R.~J. \surnamestart Duffin\surnameend} (\bibinfo{year}{1965}):
  \emph{\bibinfo{title}{Topology of Series-Parallel Networks}}.
\newblock {\sl \bibinfo{journal}{Journal of Mathematical Analysis and
  Applications}} \bibinfo{volume}{10}(\bibinfo{number}{2}), pp.
  \bibinfo{pages}{303--318}, \doi{10.1016/0022-247X(65)90125-3}.

\bibitemdeclare{inproceedings}{Habel-Plump02c}
\bibitem{Habel-Plump02c}
\bibinfo{author}{Annegret \surnamestart Habel\surnameend} \&
  \bibinfo{author}{Detlef \surnamestart Plump\surnameend}
  (\bibinfo{year}{2002}): \emph{\bibinfo{title}{Relabelling in Graph
  Transformation}}.
\newblock In: {\sl \bibinfo{booktitle}{Proc.\ International Conference on Graph
  Transformation (ICGT 2002)}}, {\sl \bibinfo{series}{Lecture Notes in Computer
  Science}} \bibinfo{volume}{2505}, \bibinfo{publisher}{Springer-Verlag}, pp.
  \bibinfo{pages}{135--147}, \doi{10.1007/3-540-45832-8\_12}.

\bibitemdeclare{inproceedings}{Manning-Plump08b}
\bibitem{Manning-Plump08b}
\bibinfo{author}{Greg \surnamestart Manning\surnameend} \&
  \bibinfo{author}{Detlef \surnamestart Plump\surnameend}
  (\bibinfo{year}{2008}): \emph{\bibinfo{title}{The {GP} Programming System}}.
\newblock In: {\sl \bibinfo{booktitle}{Proc.\ Graph Transformation and Visual
  Modelling Techniques (GT-VMT 2008)}}, {\sl \bibinfo{series}{Electronic
  Communications of the EASST}}~\bibinfo{volume}{10}.

\bibitemdeclare{article}{Plotkin04a}
\bibitem{Plotkin04a}
\bibinfo{author}{Gordon~D. \surnamestart Plotkin\surnameend}
  (\bibinfo{year}{2004}): \emph{\bibinfo{title}{A Structural Approach to
  Operational Semantics}}.
\newblock {\sl \bibinfo{journal}{Journal of Logic and Algebraic Programming}}
  \bibinfo{volume}{60--61}, pp. \bibinfo{pages}{17--139},
  \doi{10.1016/j.jlap.2004.05.001}.

\bibitemdeclare{incollection}{Plump05a}
\bibitem{Plump05a}
\bibinfo{author}{Detlef \surnamestart Plump\surnameend} (\bibinfo{year}{2005}):
  \emph{\bibinfo{title}{Confluence of Graph Transformation Revisited}}.
\newblock In \bibinfo{editor}{Aart \surnamestart Middeldorp\surnameend},
  \bibinfo{editor}{Vincent \surnamestart van Oostrom\surnameend},
  \bibinfo{editor}{Femke \surnamestart van Raamsdonk\surnameend} \&
  \bibinfo{editor}{Roel \surnamestart de~Vrijer\surnameend}, editors: {\sl
  \bibinfo{booktitle}{Processes, Terms and Cycles: Steps on the Road to
  Infinity: Essays Dedicated to {Jan Willem Klop} on the Occasion of His 60th
  Birthday}}, {\sl \bibinfo{series}{Lecture Notes in Computer Science}}
  \bibinfo{volume}{3838}, \bibinfo{publisher}{Springer-Verlag}, pp.
  \bibinfo{pages}{280--308}, \doi{10.1007/11601548}.

\bibitemdeclare{inproceedings}{Plump09a}
\bibitem{Plump09a}
\bibinfo{author}{Detlef \surnamestart Plump\surnameend} (\bibinfo{year}{2009}):
  \emph{\bibinfo{title}{The Graph Programming Language {GP}}}.
\newblock In: {\sl \bibinfo{booktitle}{Proc.\ International Conference on
  Algebraic Informatics (CAI 2009)}}, {\sl \bibinfo{series}{Lecture Notes in
  Computer Science}} \bibinfo{volume}{5725},
  \bibinfo{publisher}{Springer-Verlag}, pp. \bibinfo{pages}{99--122},
  \doi{10.1007/978-3-642-03564-7\_6}.

\bibitemdeclare{inproceedings}{Plump-Steinert10a}
\bibitem{Plump-Steinert10a}
\bibinfo{author}{Detlef \surnamestart Plump\surnameend} \&
  \bibinfo{author}{Sandra \surnamestart Steinert\surnameend}
  (\bibinfo{year}{2010}): \emph{\bibinfo{title}{The Semantics of Graph
  Programs}}.
\newblock In: {\sl \bibinfo{booktitle}{Proc.\ Rule-Based Programming (RULE
  2009)}}, {\sl \bibinfo{series}{Electronic Proceedings in Theoretical Computer
  Science}}~\bibinfo{volume}{21}, pp. \bibinfo{pages}{27--38},
  \doi{10.4204/EPTCS.21.3}.

\bibitemdeclare{article}{Poskitt-Plump12a}
\bibitem{Poskitt-Plump12a}
\bibinfo{author}{Christopher~M. \surnamestart Poskitt\surnameend} \&
  \bibinfo{author}{Detlef \surnamestart Plump\surnameend}
  (\bibinfo{year}{2012}): \emph{\bibinfo{title}{{Hoare}-Style Verification of
  Graph Programs}}.
\newblock {\sl \bibinfo{journal}{Fundamenta Informaticae}}.
\newblock \bibinfo{note}{To appear}.

\end{thebibliography}
